 \documentclass[smallabstract,smallcaptions]{dccpaper}

\usepackage{epsfig}
\usepackage{citesort}
\usepackage{amsmath}
\usepackage{amssymb}
\usepackage{color}
\usepackage{url}

\newlength{\figurewidth}
\newlength{\smallfigurewidth}

\begin{document}

\title
{\large
\textbf{Pseudo Sequence based 2-D hierarchical reference structure for Light-Field Image Compression}
}

\author{%
Li Li$^{\ast}$, Zhu Li$^{\ast}$, Bin Li$^{\star}$, Dong Liu$^{\dag}$, and Houqiang Li$^{\dag}$\\[0.5em]
{\small\begin{minipage}{\linewidth}\begin{center}
\begin{tabular}{ccc}
$^{\ast}$University of Missouri-KC & $^{\star}$Microsoft Asia & $^{\dag}$USTC \\
5100 Rockhill Road & 5 Dan Lin Street & 443 Huangshan Road \\
Kansas City, MO 64111, USA & Beijing City, 100080, China & Hefei City, 230027, China\\
\url{{lil1,lizhu}@umkc.edu} & \url{{libin@microsoft.com}} & \url{{dongeliu, lihq}@ustc.edu.cn}
\end{tabular}
\end{center}\end{minipage}}
}

\maketitle
\thispagestyle{empty}

\begin{abstract}
In this paper, we present a novel pseudo sequence based 2-D hierarchical reference structure for light-field image compression.
In the proposed scheme, we first decompose the light-field image into multiple views and organize them into a 2-D coding structure according to the spatial coordinates of the corresponding microlens. 
Then we mainly develop three technologies to optimize the 2-D coding structure.
First, we divide all the views into four quadrants, and all the views are encoded one quadrant after another to reduce the reference buffer size as much as possible.
Inside each quadrant, all the views are encoded hierarchically to fully exploit the correlations between different views.
Second, we propose to use the distance between the current view and its reference views as the criteria for selecting better reference frames for each inter view.
Third, we propose to use the spatial relative positions between different views to achieve more accurate motion vector scaling.
The whole scheme is implemented in the reference software of High Efficiency Video Coding.
The experimental results demonstrate that the proposed novel pseudo-sequence based 2-D hierarchical structure can achieve maximum $14.2\%$ bit-rate savings compared with the state-of-the-art light-field image compression method.
\end{abstract}

\section{Introduction}
\label{Sec::Introduction}
Light-field (LF) images \cite{Marc1996}, also known as plenoptic images, contain the information of both the intensity of light in a scene and the direction of the light rays in space.
Therefore, they are actually with 4-D information, not only the spatial information (similar to the traditional 2-D image) but also the angular information (different views in 2-D space).
LF images are typically acquired with the so-called plenoptic cameras through putting a microlens array just in front of the traditional imaging sensor.
Since the LF image can render refocused 3-D content, it is naturally a very attractive solution to 3-D imaging and sensing.
Especially, in recent years, along with the emergence of several kinds of plenoptic cameras, such as Lytro \cite{Ren2005} and Raytrix \cite{Christian2012}, the LF field image is becoming a more and more hot research topic.

Whether or not the LF image will be widely used in the future mainly depends on how well various challenges will be overcome.
Currently, one urgent challenge for the LF image is its huge size. 
Since the LF image is with 4-D information, even if the spatial resolution of one view is quite small, the raw data of a LF image with hundreds of views is still very large.
For example, for a raw LF image generated by Lytro \cite{CFPICME}, its resolution can be as high as $7728 \times 5368$.
Besides the huge image size, since the LF image is generated from a microlens array, its characteristic is totally different from the general 2-D image, which makes it even more difficult to compress.

The current researches for LF image compression can be mainly divided into two kinds both making full use of the classic image/video coding standards, such as JPEG \cite{Wallace1992} or High Efficiency Video coding (HEVC) \cite{Sullivan2012}.
One kind of methods uses the typical image compression standards or intra coding method of video coding standards to compress the LF image directly.
Since the LF image can be organized as a sequence of 2-D frames similar to each other, such kind of methods always also makes use of the self-similarity between different frames.
For example, Conti \emph{et al.} \cite{Conti2016} first propose to add the self-similarity motion compensation (similar to the intra block copy in screen content coding) into the HEVC coding framework to compress the LF image.
Since the motion between different views is different from the motion of intra block copy, an efficient motion prediction scheme is also proposed to reduce the header information.
Besides, the work is also extended for bi-directional prediction \cite{ContiICME2016} and weighted prediction among multiple templates \cite{Monteiro2016} to further increase coding efficiency.
Moreover, Li \emph{et al.} \cite{Li2016} propose to extend the inter P/B frame prediction in HEVC to intra prediction to better improve the LF image compression efficiency.
However, the main concern of this kind of methods is that the correlations among different views cannot be fully utilized.

The other kind of methods is the pseudo sequence based method.
As its name implies, the pseudo sequence based method tries to decompose the LF image into multiple views and organize the multiple views into a sequence to make full use of the inter correlations among various views.
For example, Perra and Assuncao \cite{Perra2016} propose to divide the LF image into multiple tiles and organize all these tiles sequentially into a sequence.
Then the correlations among different frames are exploited using the HEVC inter prediction.
Besides, Liu \emph{et al.} \cite{Liu2016} first propose to decompose the LF image into multiple views according to the relative positions of different microlens.
Then the multiple views are organized into a sequence according to their spatial position relationship. 
However, both the coding orders and reference frame management are considered in a very coarse way thus the inter dependency among different views has not been fully exploited.
Therefore, the rate-distortion (R-D) performance achieved is far from the best.

In this paper, we first follow the method of pseudo sequence to decompose the LF image into multiple views and organize all the views into a 2-D coding structure.
Then we derive a hierarchical coding order to encode the 2-D coding structure efficiently within limited number of reference frames used.
Moreover, based on the spatial coordinates of different views in the 2-D coding structure, through selecting better reference frames according to the distance between various views and performing motion vector (MV) scaling according to the relative positions of different views, we try to further improve the coding efficiency.

This paper is organized as follows.
Section \ref{Sec::2-D coding sturcture} will give a detailed description of the proposed 2-D hierarchical coding structure with limited number of frames used.
We will explain the spatial-coordinates-based reference frame selection and motion vector scaling in details in Section \ref{Sec::distance}.
The experimental results will be shown in Section \ref{Sec::experiments}.
Section \ref{Sec::conclusion} concludes the whole paper.

\section{The proposed 2-D hierarchical structure}
\label{Sec::2-D coding sturcture}
\begin{figure}[t]
\begin{center}
\epsfig{width=3.5in,file=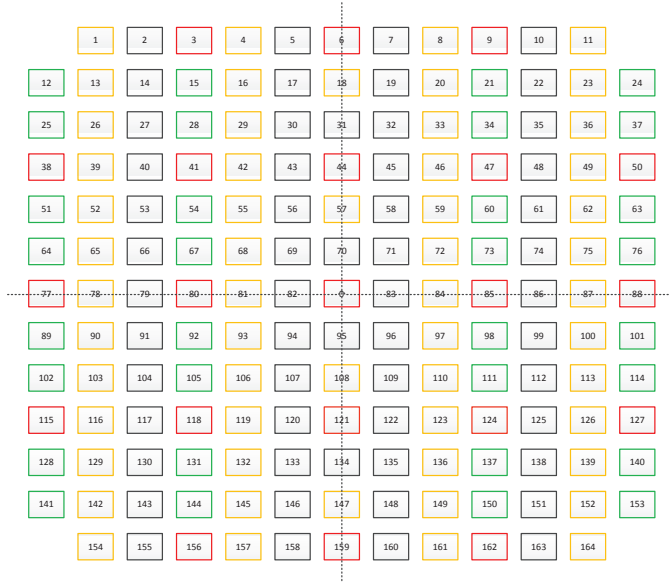}
\end{center}
\caption{\label{Fig::2-D coding structure}%
The proposed 2-D hierarchical structure}
\end{figure}

We first decompose the LF image into multiple views following the approaches in \cite{Liu2016}.
Then we delete some views in the boundary which is not beneficial for the overall LF image quality.
After these steps, for the test LF images in \cite{Rerabek2016}, we can derive $165$ views in total and assign each view to a 2-D coding structure according to the spatial position of its corresponding microlens as shown in Fig.~\ref{Fig::2-D coding structure}.
The number in each rectangular in Fig. \ref{Fig::2-D coding structure} means the picture order count (POC) assigned to each view.
Since the center view has the highest correlation with the neighboring views, the center view is assigned with POC $0$ and coded as intra frame. 
The other views are assigned the POC sequentially from top left to bottom right and coded as inter B frames.

% encoding order, reference frame selection, reference frame limitation
After assigning the POC to each frame, then comes the determination of frame coding order and reference relationships between different frames.
Here, we extend the 1-D hierarchical coding structure to construct a 2-D hierarchical coding structure to make full use of the inter correlations among all the views.
We try to simultaneously keep quite good coding efficiency and use a relatively small reference frame buffer.

As we know, in the 1-D hierarchical coding structure, the depth first coding order can lead to the least reference buffer size \cite{Schwarz2006}.
For example, The encoding order of a typical 1-D hierarchical coding structure with GOP size $16$ is $0$, $16$, $8$, $4$, $2$, $1$, $3$, $6$, $5$, $7$, $12$, $10$, $9$, $11$, $14$, $13$, $15$.
And the optimal size of the reference buffer is $5$.
The situation in the 2-D hierarchical coding structure is quite similar.
To reduce the reference buffer size as much as possible, we first divide all the frames into four quadrants as shown in Fig.~\ref{Fig::2-D coding structure}, and each quadrant will be coded in clockwise order from top left to bottom left.
In this way, except for the frames in the border of two quadrants, each quadrant can be considered as independent of each other.
Therefore, the reference frames belonging to only one quadrant can be pop out of the reference frame buffer as soon as possible.

\begin{figure}[t]
\begin{center}
\epsfig{width=2.5in,file=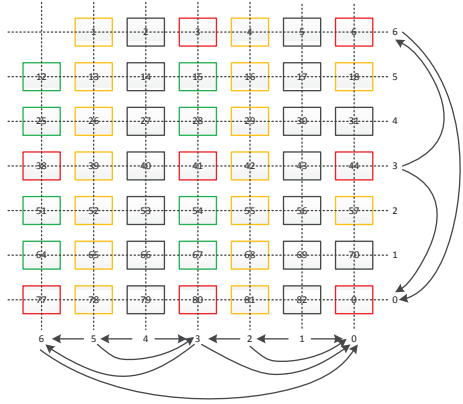}
\end{center}
\caption{\label{Fig::encoding order}%
The proposed 2-D hierarchical encoding order}
\end{figure}

Inside each quadrant, the depth first coding order will be used for both the horizontal and vertical directions.
Take the top left quadrant as an example, the detailed encoding order is shown in Fig.~\ref{Fig::encoding order}.
In both horizontal and vertical directions, we will follow the order of $0$, $6$, $3$, $5$, $4$, $2$, $1$.
And we mainly use row-based coding oder, which means that we will encode one row after another.
To be more specific, we will first encode the $0th$ row and the $0th$ column.
Then according to the hierarchical coding structure in the vertical direction, the $6th$ row will be coded and followed by the $3rd$ row.
Finally will be the $5th$, $4th$, $2nd$, and $1st$ row.
It should be also noted that to guarantee the smallest buffer size, the encoding orders of the top right and bottom right quadrants will be from right to left, and the encoding orders of the bottom right and bottom left quadrants will be from bottom to top.

After the encoding order is determined, then we will select the reference frames for each picture.
Take the top left quadrant as an example. 
As shown in Fig.~\ref{Fig::encoding order}, we first divide all the frames into four groups according to the referenced frequencies of them.
\begin{itemize}
\item The frames with the red block. The frames with the red block are the most frequently referenced frames. They are always stored in the reference buffer until they are too far away from the frames to be coded in the future. The existence of these frames can guarantee that all the frames have a relatively near reference frame.
\item The frames with the green block. The frames with the green block are the second most frequently referenced frames. They will be referenced by the frames belonging to the current row in the same quadrant. For example, besides the red frames, frame $13$ can also take frame $12$ and $15$ as references. 
\item The frames with the yellow block. The frames with the yellow block will only be referenced by the frame encoded immediately after them. For example, frame $14$ can take frame $13$ as reference.
\item The frames with the black block. The frames with the black block are the non-reference frames.
\end{itemize}
It should be noted that since we are using a row-based coding order, the vertical references are much less than the horizontal references.

Then we will analysis the size of reference buffer needed to achieve such a reference frame structure.
In the top left quadrant, according to careful analysis, we find that the maximum number of reference frames needed appears for the non-reference frames such as frame $14$.
We should not only store the eight red frames, but also frame $13$ and $15$.
Therefore, the maximum number of reference frames needed for the top left quadrant is $10$.
The situation is more complex for the top right and bottom right quadrants.
As HEVC provides the constraint that the reference frame of the next frame can only be chosen from the reference frames of the current frame and the current frame itself \cite{Sjoberg2012}, frame $77$ and $80$ should always be stored in the reference buffer because they will be used as references for the bottom left quadrant.
Therefore, the maximum number of reference frames needed for the top right and bottom right quadrants is $12$.
The situation of the bottom left quadrant is similar to that of the top left quadrant.
Therefore, in summary, the reference buffer size needed for the whole encoding process is $12$.

\section{Spatial-coordinates-based reference frame selection and MV scaling}
\label{Sec::distance}
According to the detailed analysis in the last section, we know that the number of reference frames used at most is $12$.
If we apply all these reference frames for both list0 and list1 in the encoder, the encoding complexity can be quite high.
Besides, the large reference index may also introduce extra overhead bits.
Therefore, in this section, we first propose a distance based reference frame selection to reduce both the overhead information and encoding complexity.
Then since the MV scaling process in merge \cite{Helle2012} and advanced motion vector prediction modes may also be influenced by the spatial positions of various views, we also propose a spatial-coordinates-based MV scaling to further improve the coding efficiency.

In the 1-D hierarchical coding structure, the pictures with the smaller POC difference are put in a relatively earlier position of the reference lists since they have larger possibilities to be referenced.
However, in the proposed 2-D hierarchical coding structure, the POC difference cannot reflect the distance between two frames.
For example, as shown in Fig. \ref{Fig::encoding order}, the POC difference between frame $15$ and $3$ is $12$, which is smaller than the POC difference between frame $15$ and $6$.
However, from the distance point of view, the distance between frame $15$ and $3$ is smaller than that between frame frame $15$ and $6$.
Therefore, we need to first build a relationship between the POC and the spatial position.
In our implementation, we establish a coordinate system to derive the spatial position from the POC.
The spatial coordinate of frame $0$ is set as $(0,0)$, and the left and top are set as the positive directions, the right and down are set as the negative directions.
The spatial coordinates of all the frames in the top left quadrant can be seen from Fig.~\ref{Fig::encoding order}.
According to Fig.~\ref{Fig::encoding order}, we can obviously see that the correspondence between POC and the spatial coordinates can be calculated through a look-up table or a simple formula.
After the spatial coordinates are determined, we can easily calculate the distance between frame $(x_1,y_1)$ and $(x_2,y_2)$ through Eq.~\ref{equ::distance}.
\begin{equation}
\label{equ::distance}
d = \sqrt{(x_1-x_2)^2+(y_1-y_2)^2}
\end{equation} 

Then we will construct list0 and list1 according to the distance between the current frame and its reference frames.
In the 1-D hierarchical coding structure, the forward and backward reference frames are put into list0 and list1, respectively.
Therefore, under the 2-D hierarchical coding structure, we should first define the forward and backward directions.
In our implementation, the above frames are all treated as forward frames, and the bottom frames are all treated as backward frames.
As an example shown in Fig.~\ref{Fig::encoding order}, for frame $14$, frame $1$ to frame $13$ are considered as the forward frames, and all the other frames including frame $0$ are considered as the backward frames.
The totally available reference frames for frame $14$ are $13$, $6$, $3$, $15$, $38$, $41$, $44$, $77$, $80$, and $0$.
If we set the number of reference frames in both lists as $4$, according to the distances between the current frame and its reference frames in increasing order, the reference frames in list0 will be $13$, $3$, $6$, and $15$, and the reference frames in list1 will be $15$, $41$, $38$, and $44$.
The frame $15$ is added to list0 because the number of frames in the forward is less than $4$.

Besides, the spatial coordinates may also have significant influence on the MV scaling in both merge and advanced motion vector prediction modes.
MV scaling operations are performed when the spatial neighboring blocks or temporal co-located blocks are pointing to the different reference frames from the current block.
The MV scaling can be divided into two kinds: the spatial MV scaling and the temporal MV scaling.
In 2-D hierarchical coding structure, we should perform MV scaling based on the distance in \emph{x} and \emph{y} directions instead of POC.

\begin{figure}[t]
\begin{center}
\begin{tabular}{cc}
\epsfig{width=1.5in,file=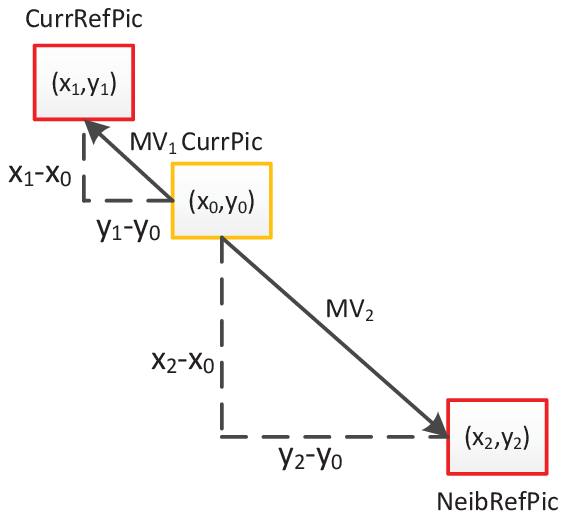} &
\epsfig{width=1.5in,file=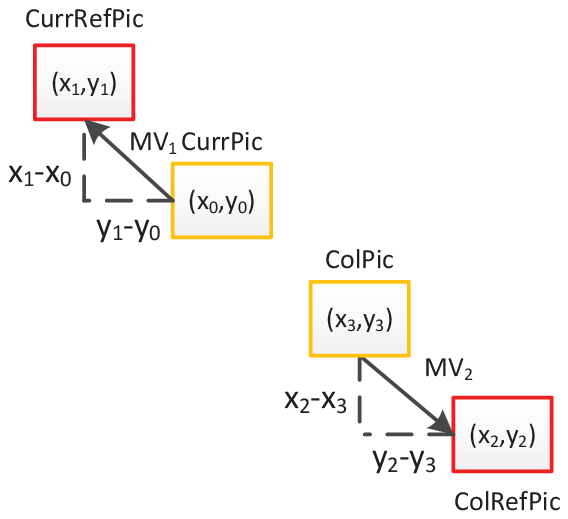} \\
{\small (a) spatial} & {\small (b) temporal}
\end{tabular}
\end{center}
\caption{\label{fig::MV scaling}%
Distance based Motion vector scaling}
\end{figure}

The detailed processes are shown in Fig.~\ref{fig::MV scaling}.
For the spatial case, the spatial coordinate of the current block is $(x_0,y_0)$, the spatial coordinate of the current reference frame is $(x_1,y_1)$, and the spatial coordinate of the reference frame of the neighboring block is $(x_2,y_2)$.
The MV of the current block is $(MV_{1,x},MV_{1,y})$, and the MV of the neighboring block is $(MV_{2,x},MV_{2,y})$.
Assuming the motions among various frames are uniform, we can derive the spatial MV scaling process as follows.
\begin{equation}
\label{spatial scaling x direction}
MV_{1,x} = \frac{MV_{2,x}}{x_2-x_0} \times (x_1-x_0)
\end{equation}

\begin{equation}
\label{spatial scaling y direction}
MV_{1,y} = \frac{MV_{2,y}}{y_2-y_0} \times (y_1-y_0)
\end{equation}
For the temporal case, except for the current picture and its reference picture, there are temporal co-located picture and its corresponding reference picture whose spatial coordinates are $(x_3,y_3)$ and $(x_2,y_2)$, respectively.
The MV of the current block is $(MV_{1,x},MV_{1,y})$, and the MV of the co-located block is $(MV_{2,x},MV_{2,y})$.
Assuming the motions among various frames are uniform, we can derive the temporal motion vector scaling process as follows.
\begin{equation}
\label{temporal scaling x direction}
MV_{1,x} = \frac{MV_{2,x}}{x_2-x_3} \times (x_1-x_0)
\end{equation}

\begin{equation}
\label{temporal scaling y direction}
MV_{1,y} = \frac{MV_{2,y}}{y_2-y_3} \times (y_1-y_0)
\end{equation}
It should be noted that the values of $x_1-x_0$ and $y_1-y_0$ are zero when the current frame and its reference frame are in the same row or column.
In this case, the scaling operations are not applied. 

\section{Experimental results}
\label{Sec::experiments}
The proposed algorithm is implemented in HM-16.7 \cite{HM16.7}.
The state-of-the-art LF image compression algorithm proposed in \cite{Liu2016} is used as the anchor to compare with the proposed method.
The BD-rate \cite{Bjontegaard2001} is used to measure the performance comparison between the proposed algorithm and the state-of-the-rate algorithm.
We use both the Y-PSNR and YUV-PSNR between the original LF image and reconstructed LF image shown in \cite{CFPICME} as an objective quality metric.
We use the LF images specified in \cite{CFPICME} as the test images.
Four intra quantization parameter (QP) points including 15, 20, 25, and 30 are tested to cover both the high bit-rate and low bit-rate cases.
The differences between inter QPs of various frames and intra QP are set the same as in \cite{Liu2016}.
We will optimize the bit allocation and QP settings for the proposed 2-D hierarchical coding structure in our future work.
The performance of the proposed algorithm is clearly analyzed in three aspects.
We will first present some experimental results on the overall framework.
Then we will analyze the benefits from the following two aspects: the proposed distance based reference frame selection and spatial-coordinates-based MV scaling.

\begin{table}[tp]
\begin{center}
\caption{\label{tab::overall performance}%
The performance of the overall framework}
{
\begin{tabular}{|c|c|c|}
\hline
test images & Y-BDrate & YUV-BDrate    \\
\hline
I01-rawLF & --7.1\% & --7.1\% \\
I02-rawLF & --5.4\% & --6.1\% \\
I03-rawLF & --5.4\% & --5.5\%  \\
I04-rawLF & --3.4\% & --2.6\%  \\
I05-rawLF & --6.5\% & --6.4\%  \\
I06-rawLF & --14.2\% & --14.2\% \\
I07-rawLF & --8.6\%  & --8.7\%  \\
I08-rawLF & --12.9\% & --12.7\% \\
I09-rawLF & --2.6\%  & --2.6\%   \\
I10-rawLF & --9.8\%  & --9.7\%  \\
I11-rawLF & --2.0\%  & --1.9\%     \\
I12-rawLF & --0.4\%  & --0.6\%     \\
Average   & --6.5\%  & --6.5\%   \\
\hline
\end{tabular}}
\end{center}
\end{table}

\begin{figure}[t]
\begin{center}
\begin{tabular}{cc}
\epsfig{width=2.5in,file=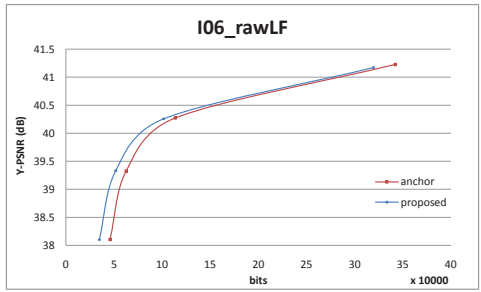} &
\epsfig{width=2.5in,file=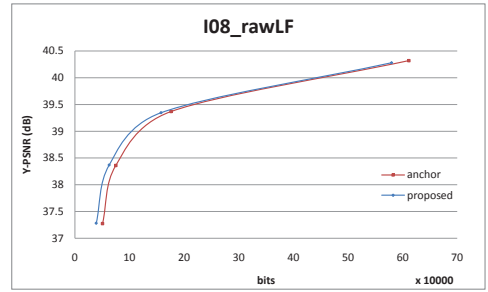} \\
{\small (a) I06 R-D vurve} & {\small (b) I08 R-D curve}
\end{tabular}
\end{center}
\caption{\label{fig:R-D curve}%
Some typical R-D curves}
\end{figure}

The performance of the overall algorithm is shown in Table~\ref{tab::overall performance}.
The results show that the proposed 2-D hierarchical reference structure can lead to significant R-D performance improvement compared with the state-of-the-art method.
In both Y-BDrate and YUV-BDrate, we can observe over $6\%$ bit-rate savings in average. 
Also we can see that the performance of the proposed algorithm is very consistent.
All the sequences present some performance improvements.
Some typical R-D curves are shown in Fig.~\ref{fig:R-D curve}.
From Fig.~\ref{fig:R-D curve}, we can obviously see that the R-D performance improvement mainly comes from the bits reduction instead of PSNR improvement.
Besides, the R-D performance improvement in low bit-rate case is much more significantly compared with the high bit-rate case.

\begin{table}[tp]
\begin{center}
\caption{\label{tab::independent performance}%
The performances of the reference framework and MV scaling}
{
\begin{tabular}{|c|c|c|c|c|}
\hline
   test         & \multicolumn{2}{c|}{reference framework}  &  \multicolumn{2}{c|}{MV scaling}            \\
images         & \multicolumn{1}{c}{Y-BDrate} & \multicolumn{1}{c|}{YUV-BDrate} & \multicolumn{1}{c}{Y-BDrate}  & \multicolumn{1}{c|}{YUV-BDrate}  \\
\hline
I01-rawLF & --6.4\% & --6.5\%  & --0.5\% & --0.4\% \\
I02-rawLF & --4.6\% & --4.6\%  & --0.6\% & --0.6\% \\
I03-rawLF & --4.7\% & --4.7\%  & --0.6\% & --0.5\% \\
I04-rawLF & --2.8\% & --2.1\%  & 0.0\% & 0.0\% \\
I05-rawLF & --5.6\% & --5.6\%  & --0.8\% & --0.7\% \\
I06-rawLF & --13.8\% & --13.7\% & --0.5\% & --0.5\% \\
I07-rawLF & --8.2\%  & --8.3\%  & --0.8\% & --0.8\% \\
I08-rawLF & --12.2\% & --12.1\% & --0.1\% & 0.0\% \\
I09-rawLF & --1.7\%  & --1.7\%  & --0.3\% & --0.3\% \\
I10-rawLF & --9.2\%  & --9.1\% & --0.1\% & --0.2\% \\
I11-rawLF & --1.1\%  & --1.0\%  & --1.9\% & --1.9\%  \\
I12-rawLF & 0.9\%    & 0.7\%    & --0.9\% & --0.9\% \\
Average   & --5.8\%  & --5.7\%  & --0.6\% & --0.6\% \\
\hline
\end{tabular}}
\end{center}
\end{table}

Both the performances of the distance-based reference frame selection and the MV scaling scheme are shown in Table~\ref{tab::independent performance}.
Through the comparison between Table~\ref{tab::overall performance} and Table~\ref{tab::independent performance}, we can see that most of the BD-rate reduction comes from the proposed distance based reference frame selection.
Since most of the MVs between different views are very small, the performance improvement brought by the proposed spatial-coordinate-based MV scaling method is quite limited.
Also, it can be obviously seen that the R-D performance improvements provided by the two parts individually are addable and can together form a better R-D performance.

\section{Conclusion and Future work}
\label{Sec::conclusion}
In this paper, we propose a pseudo sequence based 2-D hierarchical coding structure for the light-field image compression.
We first propose a 2-D hierarchical reference structure to better characterize the inter correlations among the various views decomposed from the light-field image.
Then the distance based reference frame selection and spatial-coordinate-based motion vector scaling are proposed to better utilize the 2-D hierarchical reference structure.
The proposed algorithm is implemented in the High Efficiency Video Coding reference software and quite significant R-D performance improvement is achieved compared with the state-of-the-art algorithm.
It should be noted that the bit allocation step has not been optimized for the current reference structure.
We will address the optimal bit allocation problem in our future work. 

\section{References}
\bibliographystyle{IEEEtran}{}
\bibliography{dcc}

\begin{thebibliography}{10}
\providecommand{\url}[1]{#1}
\csname url@samestyle\endcsname
\providecommand{\newblock}{\relax}
\providecommand{\bibinfo}[2]{#2}
\providecommand{\BIBentrySTDinterwordspacing}{\spaceskip=0pt\relax}
\providecommand{\BIBentryALTinterwordstretchfactor}{4}
\providecommand{\BIBentryALTinterwordspacing}{\spaceskip=\fontdimen2\font plus
\BIBentryALTinterwordstretchfactor\fontdimen3\font minus
  \fontdimen4\font\relax}
\providecommand{\BIBforeignlanguage}[2]{{%
\expandafter\ifx\csname l@#1\endcsname\relax
\typeout{** WARNING: IEEEtran.bst: No hyphenation pattern has been}%
\typeout{** loaded for the language `#1'. Using the pattern for}%
\typeout{** the default language instead.}%
\else
\language=\csname l@#1\endcsname
\fi
#2}}
\providecommand{\BIBdecl}{\relax}
\BIBdecl

\bibitem{Marc1996}
\BIBentryALTinterwordspacing
M.~Levoy and P.~Hanrahan, ``Light field rendering,'' in \emph{Proceedings of
  the 23rd Annual Conference on Computer Graphics and Interactive Techniques},
  ser. SIGGRAPH '96.\hskip 1em plus 0.5em minus 0.4em\relax New York, NY, USA:
  ACM, 1996, pp. 31--42. [Online]. Available:
  \url{http://doi.acm.org/10.1145/237170.237199}
\BIBentrySTDinterwordspacing

\bibitem{Ren2005}
\BIBentryALTinterwordspacing
R.~Ng, M.~Levoy, M.~Br\'{e}dif, G.~Duval, M.~Horowitz, and P.~Hanrahan,
  ``{Light Field Photography with a Hand-Held Plenoptic Camera},'' Tech. Rep.,
  Apr. 2005. [Online]. Available:
  \url{http://graphics.stanford.edu/papers/lfcamera/}
\BIBentrySTDinterwordspacing

\bibitem{Christian2012}
\BIBentryALTinterwordspacing
C.~Perwass and L.~Wietzke, ``Single lens 3d-camera with extended
  depth-of-field,'' pp. 829\,108--829\,108--15, 2012. [Online]. Available:
  \url{http://dx.doi.org/10.1117/12.909882}
\BIBentrySTDinterwordspacing

\bibitem{CFPICME}
M.~Rerabek, T.~Bruylants, T.~Ebrahimi, F.~Pereira, and P.~Schelkens, ``Icme
  2016 grand challenges: Light-field image compression,'' in \emph{2016 IEEE
  International Conference on Multimedia Expo Workshops (ICMEW)}, July 2016.

\bibitem{Wallace1992}
G.~K. Wallace, ``The jpeg still picture compression standard,'' \emph{IEEE
  Transactions on Consumer Electronics}, vol.~38, no.~1, pp. xviii--xxxiv, Feb
  1992.

\bibitem{Sullivan2012}
G.~J. Sullivan, J.~R. Ohm, W.~J. Han, and T.~Wiegand, ``Overview of the high
  efficiency video coding (hevc) standard,'' \emph{IEEE Transactions on
  Circuits and Systems for Video Technology}, vol.~22, no.~12, pp. 1649--1668,
  Dec 2012.

\bibitem{Conti2016}
C.~Conti, L.~D. Soares, and P.~Nunes, ``Hevc-based 3d holoscopic video coding
  using self-similarity compensated prediction,'' \emph{Signal Processing:
  Image Communication}, vol.~42, pp. 59--78, March 2016.

\bibitem{ContiICME2016}
C.~Conti, P.~Nunes, and L.~D. Soares, ``Hevc-based light field image coding
  with bi-predicted self-similarity compensation,'' in \emph{IEEE International
  Conf. on Multimedia and Expo - ICME}, July 2016, pp. 1-- 4.

\bibitem{Monteiro2016}
R.~Monteiro, L.~Lucas, C.~Conti, P.~Nunes, N.~M.~M. Rodrigues, S.~Faria,
  C.~Pagliari, E.~Silva, and L.~D. Soares, ``Light field hevc-based image
  coding using locally linear embedding and self-similarity compensated
  prediction,'' in \emph{IEEE International Conf. on Multimedia and Expo -
  ICME}, vol.~-, July 2016, pp. ---.

\bibitem{Li2016}
Y.~Li, M.~Sjöström, R.~Olsson, and U.~Jennehag, ``Coding of focused plenoptic
  contents by displacement intra prediction,'' \emph{IEEE Transactions on
  Circuits and Systems for Video Technology}, vol.~26, no.~7, pp. 1308--1319,
  July 2016.

\bibitem{Perra2016}
C.~Perra and P.~Assuncao, ``High efficiency coding of light field images based
  on tiling and pseudo-temporal data arrangement,'' in \emph{2016 IEEE
  International Conference on Multimedia Expo Workshops (ICMEW)}, July 2016,
  pp. 1--4.

\bibitem{Liu2016}
D.~Liu, L.~Wang, L.~Li, Z.~Xiong, F.~Wu, and W.~Zeng, ``Pseudo-sequence-based
  light field image compression,'' in \emph{2016 IEEE International Conference
  on Multimedia Expo Workshops (ICMEW)}, July 2016, pp. 1--4.

\bibitem{Rerabek2016}
M.~Rerabek and T.~Ebrahimi, ``New light field image dataset,'' in \emph{8th
  International Conference on Quality of Multimedia Experience (QoMEX)}, 2016.

\bibitem{Schwarz2006}
H.~Schwarz, D.~Marpe, and T.~Wiegand, ``Analysis of hierarchical b pictures and
  mctf,'' in \emph{2006 IEEE International Conference on Multimedia and Expo},
  July 2006, pp. 1929--1932.

\bibitem{Sjoberg2012}
R.~Sjoberg, Y.~Chen, A.~Fujibayashi, M.~M. Hannuksela, J.~Samuelsson, T.~K.
  Tan, Y.~K. Wang, and S.~Wenger, ``Overview of hevc high-level syntax and
  reference picture management,'' \emph{IEEE Transactions on Circuits and
  Systems for Video Technology}, vol.~22, no.~12, pp. 1858--1870, Dec 2012.

\bibitem{Helle2012}
P.~Helle, S.~Oudin, B.~Bross, D.~Marpe, M.~O. Bici, K.~Ugur, J.~Jung, G.~Clare,
  and T.~Wiegand, ``Block merging for quadtree-based partitioning in hevc,''
  \emph{IEEE Transactions on Circuits and Systems for Video Technology},
  vol.~22, no.~12, pp. 1720--1731, Dec 2012.

\bibitem{HM16.7}
\BIBentryALTinterwordspacing
{High Efficiency Video Coding test model, HM-16.7}. [Online]. Available:
  \url{https://hevc.hhi.fraunhofer.de/svn/svn_HEVCSoftware/tags/}
\BIBentrySTDinterwordspacing

\bibitem{Bjontegaard2001}
G.~Bjontegaard, ``{Calculation of average PSNR differences between
  RD-curves},'' Document VCEG-M33, Austin, Texas, USA, April 2001.

\end{thebibliography}

\end{document}